\documentclass[letterpaper]{article} 
\usepackage{aaai24}  
\usepackage{times}  
\usepackage{helvet}  
\usepackage{courier}  
\usepackage[hyphens]{url}  
\usepackage{graphicx} 
\usepackage{natbib}  
\usepackage{caption} 
\usepackage{algorithm}
\usepackage{algorithmic}

\usepackage{newfloat}
\usepackage{listings}
\DeclareCaptionStyle{ruled}{labelfont=normalfont,labelsep=colon,strut=off} 
\floatstyle{ruled}
\newfloat{listing}{tb}{lst}{}
\floatname{listing}{Listing}
\usepackage{amsmath, amsthm, amsfonts}
\usepackage{dsfont}

\newtheorem{definition}{Definition}
\usepackage{color}

\def\Eq{Eq.~}

\title{Differentially Private Training of Mixture of Experts Models\thanks{Preliminary work presented as a poster at the 5th AAAI Workshop on Privacy-Preserving Artificial Intelligence (PPAI '24).}
}
\author{
    Pierre Tholoniat\textsuperscript{\rm 1}\thanks{Work done during an internship at Microsoft Research.},
    Huseyin A. Inan\textsuperscript{\rm 2},
    Janardhan Kulkarni\textsuperscript{\rm 3},
    Robert Sim\textsuperscript{\rm 2}
}
\affiliations{
    \textsuperscript{\rm 1}Columbia University,
    \textsuperscript{\rm 2}M365 Research,
    \textsuperscript{\rm 3}Microsoft Research,\\
    
    pierre@cs.columbia.edu, \{huinan,jakul,rsim\}@microsoft.com
}

\begin{document}

\maketitle

\begin{abstract}
    This position paper investigates the integration of Differential Privacy (DP) in the training of Mixture of Experts (MoE) models within the field of natural language processing. 
As Large Language Models (LLMs) scale to billions of parameters, leveraging expansive datasets, they exhibit enhanced linguistic capabilities and emergent abilities. 
However, this growth raises significant computational and privacy concerns. 
Our study addresses these issues by exploring the potential of MoE models, known for their computational efficiency, and the application of DP, a standard for privacy preservation. 
We present the first known attempt to train MoE models under the constraints of DP, addressing the unique challenges posed by their architecture and the complexities of DP integration. 
Our initial experimental studies demonstrate that MoE models can be effectively trained with DP, achieving performance that is competitive with their non-private counterparts.
This initial study aims to provide valuable insights and ignite further research in the domain of privacy-preserving MoE models, softly laying the groundwork for prospective developments in this evolving field.
\end{abstract}

\section{Introduction}

\label{sec:intro}
The field of natural language processing has witnessed a remarkable trajectory in the development of Large Language Models (LLMs), with models continuously scaling in size and complexity. 
Rooted in the foundational principles laid out by the celebrated transformers architecture \cite{VaswaniSPUJGKP17}, LLMs today have grown to encompass billions, if not trillions, of parameters \cite{BrownMRSK20, raffel2019exploring, anil2023palm, openai2023gpt4, touvron2023llama}. 
This scaling trend is correlated with ever-expanding training datasets, as LLMs often use huge parts of the internet to capture diverse linguistic patterns \cite{kaplan2020scaling}. 
Such expansive training regimes have resulted in LLMs that not only excel in a wide array of linguistic tasks but also exhibit emergent abilities ranging from rudimentary reasoning to nuanced understanding \cite{wei2022emergent}.

In the pursuit of leveraging the benefits of increasingly large models, the computational expense of training these giants has become a significant concern, underscoring the importance of architectural efficiency. 
A particularly promising direction in this realm is the Mixture of Experts (MoE) models. 
Introduced by \cite{Jacobs91, Jordan93}, and later popularized in the context of deep learning by \cite{eigen2014learning, shazeer2017outrageously}, MoE models divide the responsibility of a neural network among several specialized sub-networks or ``experts". 
Each expert specializes in a subset of the data or task, enabling the model to allocate computation more judiciously. 
As a result, MoE architectures can achieve comparable or even superior performance to dense models for a fraction of the computational cost. 
This efficiency has made MoE models an attractive option for training extremely large models, bridging the gap between the desire for model scale and the practical constraints of training resources. 
Recent works, such as the ``Switch Transformer" \cite{Fedus22}, have showcased the potential of MoE models in scaling up while maintaining  computational footprint.

As the capabilities of LLMs continue to expand, so too do the concerns surrounding privacy. 
There is a significant risk that these immense models, if trained on private datasets, could memorize and inadvertently regurgitate private or sensitive data.
Numerous studies in the literature have demonstrated that these privacy concerns are not merely theoretical but can manifest in real-world scenarios \cite{li2023privacy, smith2023identifying}. 
In this context, Differential Privacy (DP) \cite{DworkMNS06} is considered as the gold standard for addressing these concerns by offering rigorous and quantifiable privacy guarantees, ensuring that the model remains nearly identical whether or not a specific individual's data was used during training. 
In this work, we primarily focus and initiate a study on the integration and training of MoE models with DP.
We ask:

{\em How can we effectively enable DP training for MoE models, and what implications might this have for their performance?}

\subsection{Our Contributions}
We initiate the study of training MoE models while satisfying the strong mathematical guarantees of DP \cite{dwork2014algorithmic}.
Our main contributions are:
\begin{enumerate}
    \item To the best of our knowledge, we are the first to study DP training of MoE models. 
     We identify and tackle significant challenges arising from the MoE architecture when integrated with DP optimization, especially those arising from computing the per-sample gradients in DPSGD, and present practical solutions to overcome them. 
    \item We do an empirical evaluation of our DP modifications for training of MoE models. 
    Following previous works, we consider the popular fine-tuning setting \cite{yu2022differentially} where we start with pretrained LLMs and finetune on the private dataset. 
    Through a first-round of experiments and evaluations, we show that MoE models can be effectively trained with DP and achieve competitive performance with respect to their non-private counterparts.
\end{enumerate}

Our work provides a blueprint for researchers and practitioners aiming to harness the power of MoE models under robust privacy guarantees, marking an important step forward in combining advanced model architectures and privacy-preserving methodologies.

\section{Preliminaries}
\subsection{Differential Privacy}
We formally state the definition of Differential Privacy.

\begin{definition}[Differential Privacy (DP) \citep{DworkKMMN06}]
A randomized algorithm $\mathcal{A}$ is  ($\epsilon$,$\delta$)-differentially private if for any two neighboring inputs $D$ and $D'$, which differ in only a single record, and for any set $\mathcal{S}$ of possible outputs: 
$$
\textstyle{\Pr[\mathcal{A}(D) \in \mathcal{S}] \leq e^{\epsilon}\,\Pr[\mathcal{A}(D') \in \mathcal{S}] +\delta}.
$$
\end{definition}

In the context of machine learning, Differentially Private Stochastic Gradient Descent (DPSGD) \cite{AbadiCGMMTZ16} is the most widespread way of training DP models. DPSGD adapts the conventional Stochastic Gradient Descent by implementing a per-sample gradient clipping mechanism, alongside the injection of calibrated noise into the gradient updates at every training iteration.
Furthermore, recent work \cite{li2022large, yu2022differentially} has shown impressive performance of LLMs fine-tuned with DPSGD, often being competitive to their non-private counterparts with strong privacy guarantees. In this paper, we study and demonstrate how to train MoE models with DPSGD. 

\subsection{Mixture of Experts}
\citet{Jacobs91} first propose mixture-of-experts (MoE) models, where a gating network routes an input to one of many submodels.
\citet{eigen2014learning} propose to stack multiple MoE models, thereby allowing an exponential number of effective experts.
\citet{shazeer2017outrageously} define the sparsely-gated MoE layer for transformers, where the gating network routes each token to the $k$ most relevant experts, and returns the weighted average of the expert outputs.
\citet{Fedus22} simplify this architecture by taking $k=1$.
The resulting model, called {\em Switch Transformer}, shows impressive results across a variety of tasks, including multilingual translation. In this work, we follow the formalism from the Switch Transformer, but our techniques (and some preliminary experiments) extend to other architectures such as \citet{artetxe-etal-2022-efficient}, which uses top-2 routing.

More precisely, we consider a transformer architecture and replace certain dense feed forward network (FFN) blocks by switch layers. A switch layer is defined as follows. It contains $N$ FFNs $E_1, \dots, E_N$.
Each token $x$ is passed to the gating network, which is a linear layer defined by $h(\cdot)$. 
We note $p_i(x) := \frac{\exp{h(x)_i}}{\sum_{j=1}^N \exp{h(x)_j}}$ the routing probability for expert $i \in [N]$, and $i_0 := \arg\max_{j \in [N]} h(x)_j$ the top expert.
The output of the switch layer is finally $p_{i_0}(x)E_{i_0}(x)$. 

MoE models can suffer from {\em expert collapse}, where a well-trained expert giving low loss is prioritized by the gating network, thereby receiving even more training samples at the detriment of other experts which remain forever untrained.
\citet{Fedus22} therefore add an auxiliary loss to the training loss, to help balance load across experts.
For a batch $\mathcal{B}$ with total number of tokens $T$ and a hyperparameter $\alpha >0$, this auxiliary loss is defined by:

\begin{align}
    \label{eq:batch_balancing}
    \alpha N \cdot \sum_{i=1}^{N} \frac{1}{T}\left(\sum_{x \in \mathcal{B}} \mathds{1}(\arg\max p(x) = i)\right) \cdot \frac{1}{T}\left(\sum_{x \in \mathcal{B}} p_i(x)\right)
\end{align}

where the first factor is the fraction of tokens routed to the expert $i$, and the second part is the fraction of the router probability allocated to expert $i$.

\section{Training MoE models with DP}
We identify and address three main challenges in implementing differential privacy for MoE models, all related to the per-sample gradients required by DPSGD.
Indeed, we need per-sample gradients to bound the contribution of individual samples in each batch. 
At time $t$ with model parameters $\theta_t$, the gradient for sample $s_j$ is:

\begin{equation}
g_t(s_j) = \nabla_{\theta_t} \left[ \mathcal{L}(\theta_t, s_j) + \ell(\theta_t, s_j)  \right]
\end{equation}

where $\mathcal{L}$ and $\ell$ are per-sample versions of the training and load-balancing loss, respectively.

\subsection{Per-sample balancing loss}
The first problem is that per-sample balancing loss $\ell(s_i)$ is ill-defined; note that balancing loss is defined to ensure that routing network load-balances the tokens equally among the experts.  
We only have a per-batch balancing loss $\ell(\mathcal{B})$ (defined in \Eq \ref{eq:batch_balancing}), in which all the samples are entangled: there is no immediate way of defining $\ell(s_i)$ such that $\ell(\mathcal{B}) = \sum_{j=1}^B \ell(s_j)$. 
Hence, it is not obvious how to compute the per-sample gradients of the expression from \Eq \ref{eq:batch_balancing}.

The simplest solution to address this problem is to simply remove the load-balancing loss, which is the solution we adopt.
This is particularly relevant for fine-tuning use-cases, where pretrained networks start with well-balanced gating layers.
To avoid expert collapse, it is possible to additionally freeze the gating layers.
One can also modify the load-balancing loss, replacing \Eq \ref{eq:batch_balancing} by an expression that can be decomposed into per-sample load-balancing losses.
The design of such a load-balancing loss and the investigation of its effectiveness is left for future work, as it would require us to do an ablation study of different load-balancing functions even in the non-private world.

\subsection{Expert per-sample gradient computation}
The second problem is that MoE models route tokens independently of the sample they belong to.
For dense transformer layers, there are known rules that combine activations and batch gradient (exposed by auto-differentiation engines such as PyTorch \cite{paszke2017automatic}) to reconstruct per-sample gradients \cite{Goodfellow15,li2022large}.
However, no such rules exist for MoE layers. We now detail a simple approach to compute per-sample gradients for MoE layers, illustrated by Fig. \ref{fig:routing} in Appendix. 

Consider an MoE layer that takes activations $x$ of shape $[B,T,H]$ where $B$ is the number of samples in a batch, $T$ the maximum number of tokens per sample and $H$ the hidden dimension. 
For each expert $i \in [N]$, the gating layer produces a routing mask $G_i$ of shape $[B,T]$, such that $G_i[b,t]= 1$ if token $t$ of sample $b$ is routed to expert $i$, and $G_i[b,t]=0$ otherwise.
For a given expert $i$, we can denote $C_b^i$ as the number of tokens routed to that expert from sample $b \in [B]$.
The default implementation collects tokens from different samples to form a tensor $x_i$ of shape $[\sum_{b\in[B]} C_b^i,H]$.
The lack of sample information prevents us from using per-sample gradient rules. 
We can re-introduce sample information in the routing logic by adding an extra batch dimension. 
More precisely, we now pass a tensor $x_i'$ of shape $[B, \max_{b \in [B]} C_b^i,H]$ to the expert, where $x_i'[b,c]$ contains the $c$th token from sample $b$ routed to expert $i$ if such a token exists, and $x_i'[b,c] = 0$ otherwise (e.g., if $b$ didn't route any token to expert $i$).
Once we have access to a batch dimension, we can leverage per-sample gradient rules that exist in libraries such as Opacus \cite{yousefpour2022opacus}.

The drawback of this simple approach is that it comes at a cost in memory and compute. For an expert $i$, the default MoE implementation only routes $\sum_{b\in[B]} C_b^i$ tokens to the expert. With an extra batch dimension, we route $B \times \max_{b \in [B]} C_b^i$ tokens, many of which can be zero (see Fig. \ref{fig:routing} in Appendix). 

We thus propose a more efficient and more involved algorithmic alternative in Appendix, that relies on a custom per-sample gradient rule.


\subsection{Per-sample gradients with expert parallelism}
The third challenge is to implement per-sample gradient clipping when experts are distributed across different devices (e.g. GPUs or TPUs).
Efficient parallelism is a strength of MoE models. These outrageously large models are split across many devices, but since each token only uses a subset of the model parameters, MoE can be trained with minimal computation compared to dense models with the same number of parameters.
Prior work such as GShard \cite{lepikhin2020gshard} or the Switch Transformer propose multiple parallelization techniques.
A simple and efficient form of parallelism is {\em expert parallelism}, where experts live on different devices, while dense layers are replicated across layers.
This approach can be combined with {\em data parallelism}, where each device reads a different batch, and then scatters the tokens across devices to the relevant experts.

Expert parallelism requires some adaptations to the solutions described above. For the simple solution that adds a batch dimension, we can use existing per-layer clipping techniques \cite{he2022exploring}.
For the custom per-sample rule described in Appendix, we can introduce an extra communication step that synchronizes parts of the routing table across devices, as detailed in Appendix.

Finally, \citet{Fedus22} propose to shard even dense parameters across devices, with {\em model parallelism}. This idea is also popular in Fully Sharded Data Parallelism \cite{zhao2023pytorch} or ZeRO optimization \cite{rajbhandari2020zero}. However, it is unclear how to compute and clip per-sample gradients for weights that are sharded across devices, because computing the norm of the gradients requires some cross-device communication.

\section{Experiments}
\label{sec:experiments}
\subsection{Model}
In our experiments, we use the Switch Transformer architecture \cite{Fedus22} from the open-source HuggingFace implementation and in particular start with the 8-expert pretrained model\footnote{\url{https://huggingface.co/google/switch-base-8}}. 
We start with the most straightforward solutions to the three challenges of DP presented above. First, we remove the load-balancing loss, without freezing the gating layers, relying on the observation that our pretrained networks have already good gating layers that are relatively insensitive to the load-balancing loss. Second, we introduce an extra batch dimension to leverage existing per-sample gradient implementations. Third, we do not use expert parallelism. We use simple data parallelism, since our 8-expert model is small enough for our experimental setup, which employs 16 V100 GPUs each with 32GB memory.
\subsection{Datasets}
We initiate a study on the natural language understanding setting and fine-tune the switch-base-8 model on two tasks: SST-2 and MNLI from the GLUE benchmark \cite{wang-etal-2018-glue}.
The SST-2 dataset offers a binary classification task, where sentences from movie reviews have been human-annotated with their sentiment polarity. 
The dataset has 67349 sentences for training and 872 sentences for validation.
The MNLI dataset involves the task of textual entailment, where the goal is to predict whether a given hypothesis sentence logically follows from a premise sentence.
The dataset comprises 392702 sentence pairs for training and 20000 sentence pairs for validation that are annotated with textual entailment information: entailment, contradiction, or neutral.
For DP training, we consider each sentence (or sentence pair) as a separate record.
\subsection{Hyperparameters}
Choosing hyperparameters that lead to good performance differs significantly for private and non-private fine-tuning as demonstrated by prior work \cite{li2022large, yu2022differentially}.
For non-private fine-tuning, informed by prior work \cite{Fedus22}, we set the batch size to 32 and fine-tune the switch-base-8 model for 3 epochs using the AdamW optimizer \cite{loshchilov2018decoupled} with learning rate 0.0001 and weight decay 0.01.
For private fine-tuning, informed by prior work \cite{li2022large, yu2022differentially} where large batch sizes and long training runs are shown to be effective for improved performance, we set the batch size to 1024 and fine-tune the switch-base-8 model for 20 epochs using the AdamW optimizer with learning rate 0.0005 and weight decay 0.01.
We target $\epsilon=8$ and $\delta = 1/N$ where $N$ is the size of the training dataset.
We set per-sample clipping norm to be 1.0 and calculate the corresponding noise multiplier based on the hyperparameters using the PRV accountant \cite{gopi2021numerical}.

\subsection{Results}
We report the prediction accuracy on the validation set for the two tasks in Table \ref{table:main}, similar to prior work \cite{li2022large, yu2022differentially}.

\begin{table}[h]
\centering
\begin{tabular}{c|c|c}
\hline
Method & SST-2 & MNLI \\ \hline \hline
Non-private fine-tuning & 94.5 & 85.4 \\ \hline
Private fine-tuning ($\epsilon=8$) & 92.0 & 78.7 \\ \hline
\end{tabular}
\caption{Accuracy of fine-tuning for downstream tasks with switch-base-8 model (in \%). Private fine-tuning achieves comparable performance to fine-tuning non-privately for the SST-2 dataset and it exhibits a slightly larger margin when applied to the MNLI dataset.}
\label{table:main}
\end{table}

Our initial experimental studies demonstrate promising results in terms of privately fine-tuning MoE models. 
Notably, for the SST-2 dataset private fine-tuning provides a similar performance compared to non-private fine-tuning. 
On the other hand, the MNLI dataset exhibits a larger gap between the non-private and private fine-tuning performances, which underscores the potential for further improvements of private fine-tuning of MoE models.

\section{Related Work}
\label{sec:related}
The growing importance of privacy-preserving machine learning parallels the rise of more capable and complex machine learning models, especially Large Language Models (LLMs).
This is due to the fact that this escalation in model capacity brings a heightened risk of capturing and retaining all types of information from the training data, regardless of its relevance to user interactions with LLMs.  In this context, prior work including \cite{Zanella20, CarliniTWJHLRBSEOR21, inan2021training, huang-etal-2022-large, balle2022reconstructing} has demonstrated that memorization can lead to successful reconstruction of training data.
On the other hand, membership inference attacks \cite{hu2022membership}, where an adversary aims to determine whether a particular data record was used in training a model, have also been shown to be notably effective against LLMs.
LLMs have recently undergone additional refinement steps post pre-training, such as instruction fine-tuning and alignment, to enhance their performance and user experience. 
Concurrently, these processes have raised novel privacy concerns.
There is an emerging body of literature examining the privacy implications of these post-training enhancements \cite{perez2022ignore,  wan2023poisoning}.

To address the aforementioned privacy issues with LLMs, Differential Privacy (DP) \cite{DworkMNS06} has emerged to become the gold-standard technique, offering a rigorous mathematical framework to protect individual data privacy.
In this context, recent advancements have demonstrated remarkable success and impressive results across various downstream tasks in fine-tuning pre-trained LLMs using DP \cite{li2022large, yu2022differentially, he2022exploring}.
However, these works only considered dense models, leaving MoE models understudied from a DP standpoint.

In addition to the seminal work introduced above, there is a growing body of literature and industry implementations for MoE models \cite{RLY+22, DHD+22}. A variety of routing mechanisms have been proposed \cite{ZLL+22a, ZLJ+22a}, some of which, such as random routing, might be friendlier to differential privacy. Finally, MoE models have been applied to other modalities such as vision tasks \cite{LFS+22}.

\section{Conclusion and Future Work}
\label{sec:future_work}
\label{sec:conclusion}
In this paper, to the best of our knowledge we have pioneered the study of training Mixture of Experts (MoE) models with Differential Privacy (DP). 
We give detailed the challenges that emerge from the distinct architecture and inherent complexities of MoE models when coupled with DP optimization, offering potential simple solutions to these obstacles. 
Through initial experiments and evaluations, we have shown that MoE models can be effectively trained with DP, achieving competitive performance in comparison to their non-private counterparts at least on small academic benchmarks.

Our exploration has unveiled several hard and important avenues for future work.
Beyond our initial attempts, we believe that efficient implementation of DP MoE for the state-of-the-art models with trillions of parameters matching that of non-private implementation is an important open problem and needs new ideas that build on our work.
First, a potential research direction involves revising the load-balancing loss: by decoupling it into per-sample load-balancing losses, we can examine the implications of this modification, as opposed to the current approach of simply omitting the load-balancing loss.
Second, an intriguing concept to investigate is the integration of differential privacy directly into the expert selection process.
This strategy might enable finer-grained noise addition (e.g., exclusively to the experts in use, or scaled by load) compared to the current approach of adding isotropic noise to all experts.
Last, it is crucial to broaden the scope of our experimental studies. 
By incorporating a diverse array of datasets and tasks, alongside a more extensive exploration of hyperparameters, we can attain a deeper understanding of the privacy-utility tradeoff inherent in training MoE models with DP.

\bibliography{aaai24}

\newpage
\appendix
\section{Appendix}

\subsection{Per-sample gradients with extra batch dimension}

\begin{figure}[ht!]
    \caption{{\bf Routing with extra batch dimension}. Notations are introduced in the ``Expert per-sample gradient computation" section.}
    \centering
    \includegraphics[width=0.5\textwidth]{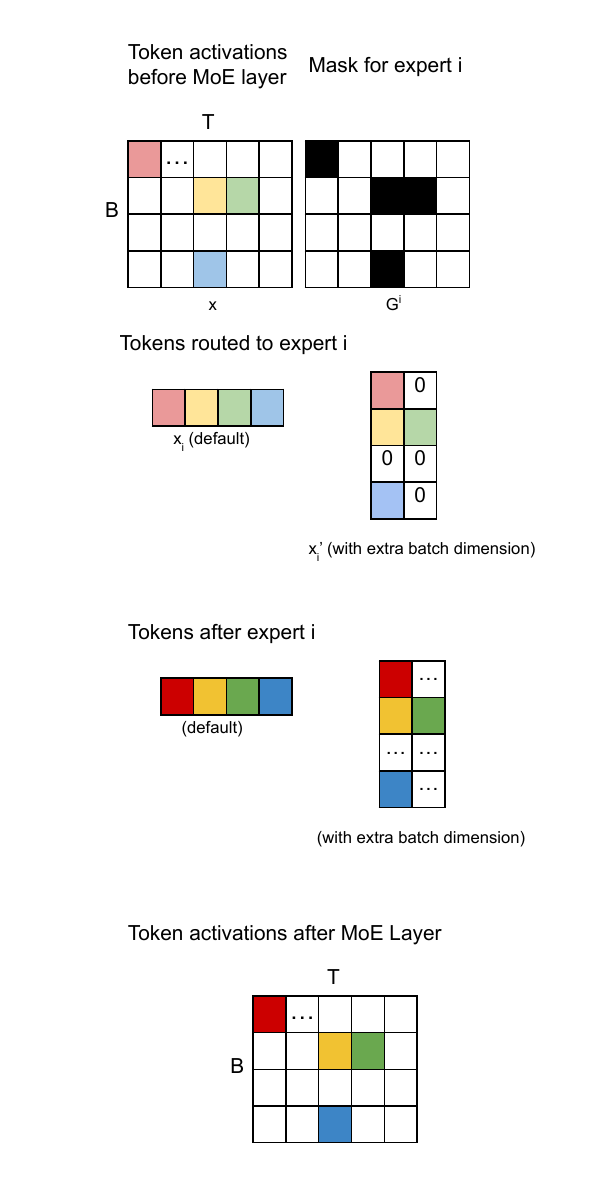}
    \label{fig:routing}
\end{figure}

\newpage
\subsection{Per-sample gradients with custom gradient rule}

In this section, we propose an alternative to the simple technique described above and illustrated in Fig. \ref{fig:routing}.
Our alternative relies on a custom gradient rule that computes per-sample gradients without additional memory cost.
The key idea is to leverage the routing table, which keeps track of the origin sample for each token.

\paragraph{Preliminary: per-sample gradients for dense layers.}

First, we restate how per-sample gradients are computed for dense linear layers.
Consider a single linear layer operating on $B$ samples containing $T$ tokens of dimension $m$, with the following notation:
\begin{itemize}
    \item Input activations: $X$ of shape $[B,T,m]$
    \item Weights: $W$ of shape $[n,m]$
    \item Output activations: $Y = XW^\top$ of shape $[B,T,n]$
    \item Model loss: $L \in \mathbb{R}$.
    \item Incoming gradients $\frac{\partial L}{\partial Y_{tk}^{(b)}}$ for $t \in [T], k \in [n], b \in B$ (e.g. as exposed by PyTorch's autograd, with a batch dimension).
\end{itemize}

A simple application of the chain rule, as in \cite{Goodfellow15}, gives that the gradients for each weight $i \in [n], j \in [m]$ can be expressed as a sum of $B$ per-sample gradients:

\begin{equation}
    \label{eq:regular_per_sample}
    \frac{\partial L}{\partial W_{ij}} = \sum_{b=1}^B \sum_{t=1}^T \frac{\partial L}{\partial Y_{ti}^{(b)}} X_{tj}^{(b)}
\end{equation}

\paragraph{Per-sample gradients for expert layers.}

Now, consider an MoE layer. Input activations $X$ of shape $[B,T,m]$ are passed to a gating network, which produces a routing table $G$  with dimensions $[E,B,T,C]$ such that: $G^{(e)}_{btc} = 1$ if token $t$ of sample $b$ is routed to slot $c$ of expert $e$, $G^{(e)}_{btc} = 0$ otherwise.

Consider a linear expert layer $e \in [E]$, with weights $W^{(e)}$ of shape $[C,m]$. It takes an input $X^{(e)}$ of shape $[C,m]$, where tokens from different samples are concatenated along a single dimension, as in Fig. \ref{fig:routing} (default case). The expert then outputs $Y = X^{(e)}(W^{(e)})^\top$ of shape $[C,n]$.

Applying Eq. \ref{eq:regular_per_sample} does not give meaningful per-sample gradients. Instead, it only gives ``per-slot'' gradients, where each sample can contribute to multiple slots:

\begin{equation}
    \label{eq:local_chain}
    \frac{\partial L}{\partial W_{ij}^{(e)}} = \sum_{c = 1}^C \frac{\partial L}{\partial Y_{ci}^{(e)}} X_{cj}^{(e)}
\end{equation}

To recover per-sample gradients, we can leverage the routing table and reassign each slot to the right sample.
Since slot $c$ of expert $e$ contains exactly one token, we have $\sum_{b=1}^{B} \sum_{t=1}^T G^{(e)}_{btc} = 1$. Hence, we can rewrite Equation \ref{eq:local_chain} as:

\begin{align}
    \frac{\partial L}{\partial W_{ij}^{(e)}} & = \sum_{c = 1}^C \left( \sum_{b=1}^{B} \sum_{t=1}^T G^{(e)}_{btc} \right) \frac{\partial L}{\partial Y_{ti}^{(b)}} X_{tj}^{(b)} \nonumber                        \\
                                             & =  \sum_{b=1}^{B} \sum_{c = 1}^C \left(\sum_{t=1}^T G^{(e)}_{btc}  \right) \frac{\partial L}{\partial Y_{ti}^{(b)}} X_{tj}^{(b)}  \label{eq:reindexed_gradients}
\end{align}

Unlike the technique from Fig. \ref{fig:routing} that added an extra dimension and zero activations in MoE routing, therefore incurring unnecessary memory and computation cost, Eq. \ref{eq:reindexed_gradients} can operate directly on the default MoE routing. It does require access to the routing table $G^{(e)}$, which is already present in memory when we operate on a single device.

\paragraph{Expert parallelism.}
To extend Eq. \ref{eq:reindexed_gradients} to multiple devices, we need to communicate the relevant parts of the routing table to each device.
It is possible to implement this efficiently, by using an all-to-all scatter operation similar to the cross-device communication already taking place to route tokens to experts. Eq. \ref{eq:reindexed_gradients} can then be adapted, with an extra dimension corresponding to the device identifier.

\end{document}